\documentclass[10pt,tightenlines,eqsecnum,floats,aps,nofootinbib,prd]{revtex4-2}
\usepackage{amsmath,amssymb,amsfonts,amsthm,amscd}
\usepackage{graphicx}
\usepackage{amsmath}
\usepackage{amssymb}
\usepackage{graphicx}
\usepackage{cancel}
\usepackage{slashed}
\usepackage{tabularx}
\usepackage{mathtools}
\usepackage{tcolorbox}
\usepackage{tikz-cd}
\usepackage{tikz}
\usepackage{mathrsfs}
\usepackage{relsize}
\usepackage{natbib}
\usepackage[margin=1in]{geometry}
\usepackage{hyperref}
\usepackage[normalem]{ulem}
\usetikzlibrary{shapes.geometric, arrows}
\tikzstyle{startstop} = [rectangle, rounded corners, minimum width=3cm, minimum height=1cm,text centered, draw=black, fill=red!30]
\tikzstyle{arrow} = [thick,->,>=stealth]
\begin{document}
\title{Constrained Symplectic and Contact Hamiltonian Systems: A Review}
\author{Callum Bell}
\email{c.bell8@lancaster.ac.uk}
\affiliation{Department of Physics, Lancaster University, Lancaster UK}

\author{David Sloan}
\email{d.sloan@lancaster.ac.uk}
\affiliation{Department of Physics, Lancaster University, Lancaster UK}

\begin{abstract}
    Singular theories, characterised by the presence of degeneracies in their Lagrangian or Hamiltonian descriptions, require the systematic implementation of constraints in order to obtain well-defined dynamics. While the symplectic framework provides the standard geometrical setting for conservative mechanical systems, those theories which exhibit dissipative effects are most appropriately discussed within the context of contact geometry. In this review, we present the geometrical structure underlying pre-symplectic and pre-contact manifolds, and develop the corresponding constraint algorithms that determine the admissible subset of phase space upon which consistent Hamiltonian evolution exists. We then close the discussion of each of the constraint algorithms with an example.
\end{abstract}
\maketitle
\section{Introduction}\label{Sec:Intro}
Singular theories of both particles and fields are amongst the most frequently encountered and physically interesting class of models in the mathematical description of natural phenomena. In the present context, we take `singular theory' to be synonymous with any model whose Hamiltonian (or Lagrangian) description must, in some way, be constrained due to the presence of degeneracies. Of course, there exist numerous kinds of singular theories, with the exact nature of the constraints being heavily context-dependent; however, of greatest interest to us will be those models which exhibit gauge invariance. Typically, such theories are addressed from the perspective of principal and/or vector bundles, in which a gauge transformation simply refers to a change of local trivialising section of the appropriate bundle. A more detailed exposition of this treatment may be found in \cite{nakahara2018geometry}, for example.\\

In general, we shall concern ourselves with singular theories of two distinct types: those described by symplectic Hamiltonians, and those with contact Hamiltonians. Symplectic geometry is by far the prevailing framework within which analytical mechanics is formalised \cite{da2008lectures}. Such manifolds are characterised as being of even dimension, and possessing a closed, non-degenerate differential 2-form $\omega$. A theorem due to Liouville guarantees that volumes of subsets of any symplectic manifold are preserved under the flow of Hamiltonian vector fields \cite{arnol1990symplectic}. As such, symplectic geometry provides a particularly elegant approach to the description of conservative mechanical systems. In section (\ref{Sec:GeometricPreliminaries}), we review a number of elementary geometrical concepts, which allow us to describe pre-symplectic Hamiltonian systems. Section (\ref{Sec:Pre-symplecticAlgorithm}) is then dedicated to a detailed treatment of the process of systematically deducing the subset of a system's phase space upon which the Hamiltonian equations of motion admit well-defined solutions. This constraint algorithm, first developed in \cite{Nester1}, uses geometrical principles, rather than the more familiar algebraic methods of the Dirac-Bergmann procedure \cite{dirac2013lectures}. For the reader acquainted with this latter method of constraint, section (\ref{Sec:Dirac-Bergmann}) provides a discussion of the similarities and differences between Dirac's construction and the geometrical ideas used throughout.\\

The second broad category of singular systems of interest are those whose phase spaces are pre-contact manifolds \cite{bravetti2017contact}. Contact geometry is far less prevalent within the physics literature. This is principally due to the fact that, unlike symplectic manifolds, the contact counterparts are odd-dimensional, with a phase space measure which is not conserved, but instead may undergo focusing/defocusing effects. By virtue of this property, contact Hamiltonian systems are frequently employed to describe dissipative phenomena. In section (\ref{Sec:ContactGeometry}), we develop the formalism of contact geometry, together with its implementation in the description of Hamiltonian systems. Section (\ref{Sec:PreContactConstraintAlgorithm}) then contains the geometrical constraint algorithm for pre-contact systems, and closely mirrors the scheme presented in section (\ref{Sec:Pre-symplecticAlgorithm}). The principal motivation for studying these non-conservative theories resides in a particular kind of symmetry reduction, known as contact reduction \cite{sloan2018dynamical}. We do not explicitly treat this procedure in the present work; however, the interested reader is invited to consult \cite{bravetti2023scaling,sloan2021scale}, and references therein.\\

Somewhat informally, it is often found that classical symplectic systems admit a vector field, whose effect is to Lie-drag the symplectic form to itself, and to rescale the Hamiltonian by some non-zero factor. Physically, this vector field - known as a dynamical similarity - is the generator of a phase space rescaling, that has the particular property of mapping between mathematically distinct, but observationally indistinguishable solutions to the equations of motion. The premise of contact reduction is then to quotient the symplectic phase space by the one-dimensional orbits of the dynamical similarity. In doing so, the phase space dimension is reduced by one, and naturally inherits a contact structure. Note that this is structurally quite distinct from the more familiar reduction schemes of Marsden and Weinstein \cite{marsden1974reduction}, in which the symmetry reduction reduces the phase space dimension by two, thereby preserving its symplectic nature.
\section{Geometrical Preliminaries}\label{Sec:GeometricPreliminaries}
We begin with a number of fundamental ideas surrounding the symplectic description of Lagrangian and Hamiltonian systems. Suppose that $Q$ is an $n$-dimensional smooth manifold, corresponding to the configuration space of just such a system. A Lagrangian function $L:TQ\rightarrow \mathbb{R}$ is introduced on the tangent bundle $TQ$, with projection $\pi:TQ\rightarrow Q$. Fibre coordinates on $TQ$ are denoted $v^i$, so that, given some $q\in U\subset Q$, induced bundle coordinates on $\pi^{-1}(U)$ are given by $(q^i,v^i)$.\\

Every tangent bundle is endowed with a ($1^{\textrm{st}}$-order)  almost-tangent structure \cite{prieto2014geometrical}, provided by the \textbf{vertical endomorphism} $J:T(TQ) \rightarrow T(TQ)$ which satisfies
\begin{itemize}
    \item[$\star$] $J^2= \textrm{id}_{TQ}$
    \item[$\star$] $\textrm{rank}\,J=n$
\end{itemize}
In order to construct this endomorphism, we briefly review the concept of the \textbf{vertical lift} of a tangent vector \cite{de2020review}. Let $X$ and $u$ be vectors in the same tangent space $T_qQ$ at $q\in Q$. The vertical lift of $X$ to $T(TQ)$ is written as
\begin{equation}
    (X)^V_u = \frac{d}{dt}\biggr|_{t=0} \left(u+tX\right)
\end{equation}
In local bundle coordinates $(q^i,v^i)$, we write 
\begin{equation*}
    X=X^i\,\frac{\partial}{\partial q^i}\biggr|_q\quad\quad\implies \quad\quad (X)^V_u = X^i\, \frac{\partial}{\partial v^i}\biggr|_u
\end{equation*}
The vertical endomorphism $J$ is then defined via its action on elements of $T(TQ)$. In particular, starting with a point $q\in Q$, we take a tangent vector $u\in T_qQ$, and consider some element $Z\in T_u(TQ)$ defined at $u$. The action of $J$ on the vector $Z$ reads 
\begin{equation}
    J(Z) = (D_{u}\pi (Z))^V_u
\end{equation}
where $D\pi:T(TQ)\rightarrow T(TQ)$ denotes the pushforward. This defines a tensor field of type $(1,1)$, and in local bundle coordinates, we have
\begin{equation}\label{Eq:VerticalEndo}
    J= \frac{\partial}{\partial v^i} \otimes dq^i
\end{equation}
In addition to the vertical endomorphism $J$, the tangent bundle $TQ$ admits a second canonical geometrical structure: the \textbf{Liouville vector field} $\Delta\in \mathfrak{X}^{\infty}(TQ)$. This is defined to be the vector field whose flow generates dilatations on the fibres of $TQ$. More specifically, if 
\begin{equation}
    \Delta_u = \frac{d}{dt}\biggr|_{t=0}\Phi_t(u)
\end{equation}
so that $\Phi:\mathbb{R}\times TQ\rightarrow TQ$ is the flow of the vector field $\Delta$, then we have 
\begin{equation}
    \Phi_t(u) = e^tu
\end{equation}
From this, it follows that $\Delta$ may be written locally as 
\begin{equation}
    \Delta= v^i\,\frac{\partial}{\partial v^i}
\end{equation}
From these structures, we introduce a 1-form $\theta_L\in \Omega^1(TQ)$
\begin{equation}
    \theta_L := \iota_J\,dL=  dL\circ J \quad\quad\implies\quad\quad \theta_L = \frac{\partial L}{\partial v^i}\,dq^i
\end{equation}
where we have used the local expression (\ref{Eq:VerticalEndo}) for the vertical endomorphism $J$. The form $\theta_L$ allows us to naturally define a closed 2-form on $TQ$ as $\omega_L := -\,d\theta_L$. The cases of greatest interest throughout will be those in which $\omega_L$ is a pre-symplectic form: that is, when it is closed, but degenerate. This may be expressed as the requirement that the matrix of second derivatives
\begin{equation*}
    W_{ij} := \left( \frac{\partial^2 L}{\partial v^i \partial v^j} \right)
\end{equation*}
be of non-maximal rank, and thus singular. Finally, a pre-symplectic Lagrangian system is a triple $(TQ,\omega_L,E_L)$, in which the scalar energy function $E_L\in C^{\infty}(TQ)$ is defined according to 
\begin{equation}
    E_L := \Delta(L) - L \quad\quad\implies\quad\quad E_L = v^i\,\frac{\partial L}{\partial v^i} - L
\end{equation}
In what follows, we shall focus almost exclusively on the Hamiltonian formalism, using the above construction of the triple $(TQ,\omega_L,E_L)$ as a means to obtain the corresponding Hamiltonian system. Our motivation for this lies principally in the fact that the Hamiltonian framework provides a well-defined bracket structure, which readily allows us to classify constraint functions, and compute the dynamical evolution of phase space variables. Such a bracket structure is notably absent on the Lagrangian side.\\

For pre-symplectic Lagrangian systems, the Legendre map $FL:TQ\rightarrow T^*Q$ is not a diffeomorphism; in practice, this translates into the more intuitive statement that one cannot `invert' the momenta to solve for the $v^i$, and write down a Hamiltonian in a straightforward fashion. In what follows, we restrict our attention to those cases in which the Lagrangian is described as being \textbf{almost-regular}; such systems are characterised as follows
\begin{itemize}
    \item[$\star$] $M_0:= FL(TQ)$ is a closed submanifold of $T^*Q$
    \item[$\star$] $FL$ is a submersion onto its image
    \item[$\star$] For every $p\in M_0$, the fibres $FL^{-1}(p)$ are connected submanifolds of $TQ$
\end{itemize}
Given an almost-regular Lagrangian system $(TQ,\omega_L,E_L)$, we denote the restriction of the Legendre map to its image via $FL_0$; since this is a submersion, it follows that there exists a \textit{unique} function $H_0:M_0\rightarrow \mathbb{R}$, such that $FL_0^*\hspace{0.3mm}H_0=E_L$. On the cotangent bundle, there exists a canonical symplectic form $\omega$, expressed in local Darboux coordinates as $\omega=dq^i\wedge dp_i$; if $\jmath_0:M_0\hookrightarrow T^*Q$ denotes the inclusion map, then $M_0$ inherits a pre-symplectic form $\omega_0:=\jmath_0^*\,\omega$, which we note also satisfies $FL_0^*\,\omega_0=\omega_L$. We shall refer to the triple $(M_0,\omega_0,H_0)$ as a pre-symplectic Hamiltonian system, defined on the \textbf{primary constraint manifold} $M_0$.\\

The dynamical problem is formulated introducing a bundle morphism $\flat:T(T^*Q)\rightarrow T^*(T^*Q)$, with $\flat(X):=\iota_X\omega$; this is then restricted to $M_0$, with the resulting map being denoted $\flat_0$. We now seek a vector field $X_H$, such that
\begin{equation}\label{Eq:HamiltonGeometric}
    \flat_0(X_H)=\iota_{X_H}\omega_0 = dH_0
\end{equation}
Solutions to this equation generally do not exist on the entirety of $M_0$, thereby necessitating the implementation of a constraint algorithm. Such a procedure gives the maximal subspace of the cotangent bundle upon which the dynamical problem (\ref{Eq:HamiltonGeometric}) possesses well-defined (albeit non-unique) solutions.
\section{The Pre-symplectic Constraint Algorithm}\label{Sec:Pre-symplecticAlgorithm}
Before reviewing how one uses geometrical principles to systematically restrict the phase space of a singular system, it will be of use to establish a number of notational conventions. Firstly, we denote by $\langle\,\cdot\,,\cdot \,\rangle$ the natural pairing between a vector space and its dual, writing, for example, $\langle dH_0,TM_0\rangle$ to refer to the contraction of a particular object, such as $dH_0\in T^*M_0$, with all elements of the space $TM_0$.\\

In general, a pre-symplectic form $\omega$ on a manifold $M$ allows us to introduce a notion of orthogonality; in particular, for any subspace $N\subset M$, we define the \textbf{symplectic orthogonal} (or symplectic complement) of $TN$, denoted $TN^{\perp}$, as follows
\begin{equation}\label{Eq:SymplecticOrthogonal}
    TN^{\perp}:=\{ X\in TM|_{\scriptscriptstyle N}\;|\; \omega|_{\scriptscriptstyle N}(X,Y)=0\;\,\textrm{for all}\; Y\in TN\,\}
\end{equation}
In addition to the symplectic orthogonal, we also introduce the \textbf{annihilator} of a subspace $S\subset TM$ as
\begin{equation}\label{Eq:Annihilator}
    S^o:=\{\alpha\in T^*M\;|\; \langle \alpha,v\rangle = 0 \;\;\textrm{for all}\; v\in S\,\}
\end{equation}
The manifold $M$ is said to be \textbf{reflexive} if $(M^*)^* = M$, and \textbf{topologically closed} if $\flat:TM\rightarrow T^*M$ maps closed sets of $TM$ onto closed sets of $T^*M$ \cite{Nester1}; for this class of space, we find the particularly useful relationship $(TN^{\perp})^o= \flat(TN)$.
\subsection{The Hamiltonian Algorithm}\label{Subsec:PresymplecticHamiltonianAlgorithm}
Let $(M_0,\omega_0,H_0)$ be the pre-symplectic Hamiltonian system, obtained from an almost-regular Lagrangian $L$. It will often be convenient to describe $M_0$ as the zero-set of a collection of \textbf{primary constraint functions} $\{\phi^a\}$, with $a=1,\,\cdots, n-k$, in which $n=\textrm{dim}\,Q$, and $k = \textrm{rank}\,W_{ij}$. We shall assume that all subspaces generated are regular, closed submanifolds of $T^*Q$, whose natural injections are embeddings, and are denoted $\jmath_i:M_i\hookrightarrow M_{i-1}$. We therefore begin by seeking a vector field $X_H$ which satisfies
\begin{equation}\label{Eq:HamiltonEqsAgain}
    \flat_0(X_H) = \iota_{X_H}\omega_0=dH_0
\end{equation}
Since $\omega_0$ is pre-symplectic, $\flat_0$ is not an isomorphism, and so we should consider only those points of $M_0$ at which $dH_0\in \textrm{Im}\,\flat_0$. From above, under the assumptions of reflexivity and topological closedness, we recall that $\flat_0(T_xM_0) = (T_x M_0^{\perp})^o$. However, $T_xM_0^{\perp}$ is precisely $\textrm{ker}\,(\omega_0)_x$, and so the subset of $M_0$ of interest is
\begin{equation}\label{Eq:M1Again}
    M_1 = \{ x\in M_0\;|\; \langle (dH_0)_x,\textrm{ker}\,(\omega_0)_x\rangle = 0\,\}
\end{equation}
In order for any vector field solution $X_H$, defined over the points of $M_1$, to be physically meaningful, it must remain \textit{tangent} to $M_1$. If this were not the case, the solution would tend to evolve off the constraint surface, and thus cease to obey the restrictions placed upon the system. Tangency of the solution is not guaranteed, and so must be imposed as an additional constraint, leading us to consider only the following subset of points of $M_1$
\begin{equation}\label{Eq:M2}
    M_2:=\{x\in M_1\;|\; \langle (dH_0)_x, T_xM_1^{\perp}\rangle =0\,\}
\end{equation}
Vector field solutions on $M_2$ are now tangent to $M_1$, but, in general, not to $M_2$, requiring this condition to be imposed as a further restriction. It is then clear how the algorithm must proceed: at the $k^{\textrm{th}}$ step, the submanifold $M_k$ is composed of all points of $M_{k-1}$ at which the solution is tangent to $M_{k-1}$, but not necessarily to $M_k$ itself. This leads us to restrict $M_k$ to the smaller set $M_{k+1}$, which may be imposed requiring that $dH_0$ belong to the annihilator $(TM_k^{\perp})^o$, so that
\begin{equation}\label{Eq:Mk+1}
    M_{k+1} = \{x\in M_k\;|\; \langle (dH_0)_x , T_xM_k^{\perp}\rangle=0\,\}
\end{equation}
The constraint procedure may terminate in one of three ways \cite{Nester1}; however, the only dynamically-interesting eventuality is that there exists some $N\in\mathbb{N}$, such that $M_{N+1}=M_N$, with $\textrm{dim}\,M_N\neq0$. Upon deducing the value of $N$ for which the algorithm terminates, we shall declare $M_f:=M_N$ to be the \textbf{final constraint submanifold}.\\

Practical implementation of this algorithm, as we shall soon see, is frequently far less convoluted than one might expect from the series of formal statements given above. Indeed, for the majority of systems of interest, the algorithm stabilises after two or three iterations, minimising the need for lengthy calculations. In section (\ref{Sec:Eg1}), we provide an example of how one uses the geometrical constraint algorithm to deduce the final submanifold. In order to be of greater practical use, after having implemented the algorithm, it is often of benefit to work locally on $M_f$; this requires a more detailed discussion of the geometry of this space, together with the introduction of a local bracket structure.
\subsection{Geometry of the Final Constraint Submanifold}\label{Subsec:Geometry}
Suppose that the constraint surface $M_f$, deduced from the algorithm above, is defined as the zero-set of the functions $\phi^{\alpha}$, with $\alpha=1,\,\cdots,D$. A quantity $\zeta$ defined on the full cotangent bundle $T^*Q$ is said to be \textbf{weakly vanishing}, expressed symbolically as $\zeta\approx 0$ if it is identically zero, when restricted to $M_f$. We group constraints into two broad categories: \textbf{first-class} constraints are those whose Poisson brackets with all other constraints are weakly vanishing. Any constraint that is not first-class is then denominated \textbf{second-class}. For the vast majority of mechanical systems, it transpires that all first-class functions generate gauge transformations. It was proposed by Dirac that this should be true in all physical cases - an afformation known as the \textbf{Dirac conjecture} - however, this is demonstrably false \cite{GRACIA1988355}, as there exist known counterexamples. In the present context, we lose no generality in taking the Dirac conjecture to be true, and so we shall speak interchangeably of first-class constraints and gauge symmetry generators.\\

The analysis of a system's dynamics is greatly facilitated by considering expressions in a local chart of coordinates. While we have implicitly introduced such coordinates in speaking of local constraint functions, it should be relatively clear that the submanifolds $M_i$ and vector field solution $X_H$ exist independently of any particular choice of local coordinates. From now on, however, we shall work exclusively in a particular set of coordinates - Darboux coordinates - and proceed to introduce the matrix of Poisson brackets $J^{\alpha\beta}:=\{\phi^{\alpha},\phi^{\beta}\}$. Locally, this Poisson bracket is expressed as
\begin{equation}\label{Eq:LocalPB}
    J^{\alpha\beta}:=\{\phi^{\alpha},\phi^{\beta}\} = \frac{\partial\phi^{\alpha}}{\partial q^i}\frac{\partial\phi^{\beta}}{\partial p_i} - \frac{\partial\phi^{\alpha}}{\partial p_i}\frac{\partial\phi^{\beta}}{\partial q^i}
\end{equation}
If $\textrm{rank}\,J=D-F$, it follows that there exist $F$ $C^{\infty}$-linear combinations of the $\phi^{\alpha}$ which correspond to first-class constraints, and hence to generators of gauge transformations. The goal is then to identify such combinations, and to modify the original set of constraints, so as to extract a maximal subfamily of second-class functions. To this end, we introduce a `vector' $v_{\alpha}=(v_1,\,\cdots,v_D)$, and demand that 
\begin{equation}\label{Eq:FirstClassv}
    v_{\alpha}J^{\alpha\beta}\approx  0\quad \textrm{for all}\;\beta=1,\,\cdots,D
\end{equation}
Having identified all such linearly independent combinations, indexing the resulting vectors via $v^{(a)}$, the $a^{\textrm{th}}$ first-class constraint is simply $\Omega^a:=v^{(a)}_{\alpha}\phi^{\alpha}$. In order to extract the maximal subfamily of second-class constraints, we seek $D-F$ linearly independent vectors $w^{(j)}$, which are \textit{not} in the kernel of $J$, so that
\begin{equation*}
    \underbrace{\Omega^a=v^{(a)}_{\alpha}\phi^{\alpha}}_{1^{\textrm{st}}\;\textrm{class}}\quad\quad\quad\quad\quad \underbrace{\chi^{j}=w^{(j)}_{\alpha}\phi^{\alpha}}_{2^{\textrm{nd}}\;\textrm{class}}
\end{equation*}
The constraint surface $M_f$ may now be characterised as
\begin{equation}\label{Eq:FinalM_f}
    M_f = \{\,x\in T^*Q\;|\; \Omega^a(x)=0 \quad \textrm{for}\;a=1\,,\cdots,F \quad\textrm{and} \quad\chi^j(x)=0\quad\textrm{for}\; j=1,\,\cdots,D-F\;\}
\end{equation}
In the presence of second-class functions, the usual Poisson bracket is replaced with the Dirac structure $\{\,\cdot\;, \,\cdot\,\}_D$ \cite{ibort1999dirac}. The matrix $C^{ij}:=\{\chi^i,\chi^j\}$ is, by construction, of maximal rank $D-F$, and thus non-singular. If $C_{ij}$ are the elements of the inverse $C^{-1}$, the Dirac bracket of two phase space functions $f,g\in C^{\infty}(T^*Q)$ is given by
\begin{equation}\label{Eq:DiracBracket}
    \{f,g\}_{D}:=\{f,g\} - \{f,\chi^i\} \,C_{ij}\{\chi^j,g\}
\end{equation}
Rather than restricting directly to $M_f$, as in (\ref{Eq:FinalM_f}), let us first impose the second-class constraints $\chi^j=0$ as \textit{strong equalities}. This gives a hypersurface
\begin{equation}\label{Eq:MChi}
    M_{\chi}:=\{x\in T^*Q\;|\; \chi^j(x) = 0\quad\textrm{for}\; j=1,\,\cdots, D-F\,\}\;\subset \;T^*Q
\end{equation}
that inherits a closed 2-form $\omega_{\chi}$ which is \textit{symplectic}. The Dirac bracket is precisely the algebraic medium required to `transfer' the symplectic structure of $M_{\chi}$ to functions defined on the whole of $T^*Q$. By this, we mean that $\{\,\cdot\;, \,\cdot\,\}_D$ is a modification of the standard Poisson bracket, which allows the second-class constraints to be imposed as strong equalities before computing the equations of motion. Ordinarily, in order to correctly describe a system's dynamical evolution, one must compute the desired Poisson brackets, retaining all constraints as weak equalities. Only \textit{after} may one impose the second-class constraints as strong equalities. The virtue of using the Dirac bracket is that the ability to impose the second-class constraints as strong equalities \textit{before} computing the equations of motion often simplifies the algebra enormously.\\

We now introduce a function $H$, which is any extension of $H_0$ to $T^*Q$, such that $H|_{M_0}=H_0$, and construct the \textbf{total Hamiltonian} $H_T:=H+\lambda_a\Omega^a$, in which $\lambda^a$ are undetermined Lagrange multipliers. The most general expression for the evolution of any phase space function $f\in C^{\infty}(T^*Q)$, including those trajectories related by gauge symmetries, may be deduced from $\dot{f}=\{f,H_T\}_D$.\\ 

Imposing the second-class constraints as strong equalities is equivalent to situating ourselves on the hypersurface $M_{\chi}$. Within this space, we consider the first class surface
\begin{equation}\label{Eq:MOmega}
    M_{\scriptscriptstyle\Omega}:=\{x\in M_{\chi}\;|\; \Omega^a(x)=0\quad\textrm{for}\;a=1,\,\cdots,F\,\}
\end{equation}
$M_{\scriptscriptstyle\Omega}$ is an embedded submanifold of $M_{\chi}$, whose 2-form $\omega_{\scriptscriptstyle\Omega}$ is maximally degenerate. Further, at each point $x\in M_{\scriptscriptstyle\Omega}$, the symplectic orthogonal $T_xM_{\scriptscriptstyle\Omega}^{\perp}$ satisfies $T_xM_{\scriptscriptstyle\Omega}^{\perp}\subseteq T_xM_{\scriptscriptstyle\Omega}$, and so $M_{\scriptscriptstyle\Omega}$ constitutes a coisotropic submanifold of $M_{\chi}$. For such a class of space, the distribution $x\mapsto T_xM_{\scriptscriptstyle\Omega}^{\perp}$ is involutive, and may locally be written as the span of the Hamiltonian vector fields $X_a$ corresponding to $\Omega^a$ via $\iota_{X_a}\omega_{\chi}=d\Omega^a$. That is to say, we have
\begin{equation*}
    T_xM_{\scriptscriptstyle\Omega}^{\perp} = \textrm{span}(X_1|_x\,,\cdots,X_F|_x) \,\subseteq\, T_xM_{\scriptscriptstyle\Omega}
\end{equation*}
Since the Hamiltonian vector fields of the first-class constraints define an involutive distribution, we know that there exists a foliation of $M_{\scriptscriptstyle\Omega}$ into $F$-dimensional gauge leaves. The gauge orbits correspond to integral curves of the $X_a$, and they map points of $M_{\scriptscriptstyle\Omega}$ into other points, which produce dynamically indistinguishable configurations. Quotienting out by these gauge orbits, we obtain a physical phase space $\mathcal{P}$, which inherits a closed 2-form $\omega_{\scriptscriptstyle\mathcal{P}}$ that is \textit{symplectic}.
\section{The Dirac-Bergmann Procedure}\label{Sec:Dirac-Bergmann}
Throughout, we have made reference on a number of occasions to the foundational work \cite{dirac2013lectures} of Dirac on the theory of constrained Hamiltonian systems. Indeed, \cite{dirac2013lectures} is a highly readable account, and the interested reader is encouraged to consult this reference. Here, we would like to provide a brief comparison between the geometrical algorithm described in the present article, and that of Dirac. Of course, the final dynamical content of the two procedures is identical; however, the approach presented here has the advantage that the geometrical principles used to deduce the final constraint submanifold may be extended to somewhat more general contexts.\\

In a similar fashion to our geometrical procedure, the Dirac-Bergmann algorithm also requires that we obtain the Hamiltonian $H_0$ as a naïve Legendre transform of the singular Lagrangian. In this context, $H_0$ is often referred to as the \textbf{canonical Hamiltonian}. In applying the Legendre map to $L$, we obtain a number $P$ of primary constraints $\phi^{\alpha}$, $\alpha=1,\,\cdots ,P$. The first step of the constraint process is then to add to $H_0$ the linear combination $\lambda_{\alpha}\phi^{\alpha}$ of the primary constraints. At present, the $\lambda_{\alpha}$ are unknown Lagrange multipliers. We thus define the \textbf{primary Hamiltonian} $H_P$ as
\begin{equation}\label{Eq:PrimaryH}
    H_P:=H_0+\lambda_{\alpha}\phi^{\alpha}
\end{equation}
The primary constraints we have found must hold at all times, else our constrained system would tend to evolve towards an unrestricted configuration. The weak vanishing of the $\phi^\alpha$ defines the primary constraint surface; if these constraints are to be preserved under the Hamiltonian flow, then we require that $\{\phi^\alpha,H_P\}\approx0$ for all $\alpha=1,\,\cdots,P$. In general, each of these conditions leads to one of four distinct possibilities. The two most uninteresting of these are either that $\{\phi^\alpha,H_P\}\approx0$ produces a statement such as $1=0$, which is manifestly false. Such inconsistencies imply that the system under study possesses no admissible dynamics. Alternatively, $\{\phi^\alpha,H_P\}$ could identically vanish on the constraint surface, giving a trivial expression of the form $0=0$.\\

One of the more dynamically interesting eventualities is that imposing $\{\phi^{\alpha},H_P\}\approx0$ yields an algebraic expression involving the Lagrange multipliers. Such relationships fix one of these Lagrange multipliers in terms of phase space variables, and possibly other multipliers. Finally, if the algebraic expression that arises from the weakly-vanishing Poisson bracket contains no undetermined Lagrange multipliers, then we have identified a \textbf{secondary constraint}.\\

Note that the principle that led us to impose that $\{\phi^\alpha,H_P\}\approx0$ is analogous to that which motivated us to restrict the primary submanifold $M_0$ to the subset $M_1$, consisting of only those points at which the vector field solution $X_H$ was tangent to $M_0$. In both cases, this step was carried out so as to ensure that the primary constraints were preserved under the dynamical evolution of the system. The geometrical picture is somewhat more intuitive, as it allows us to see that, were the vector field $X_H$ \textit{not} tangent to $M_0$, the dynamics of the system would tend to evolve off the constraint surface, rather than being confined to reside within it.\\

In order to continue with the Dirac-Bergmann constraint procedure, each secondary constraint $\psi^r$ must itself be subjected to the requirement that $\{\psi_r,H_P\}\approx0$. In doing so, we obtain one of the four possible outcomes described above. This process should be repeated recursively until no new constraints are obtained, and no new information is acquired about the Lagrange multipliers. We note that one may be tempted to consider adding the subsequently generated constraints to the primary Hamiltonian (\ref{Eq:PrimaryH}), with their own Lagrange multipliers. This is, in general, not the correct way to proceed. The origin of the undetermined Lagrange multipliers in (\ref{Eq:PrimaryH}) lies in the uncertainty/ambiguity that arises when passing from the singular Lagrangian to the canonical Hamiltonian. More mathematically, the Legendre map is not a diffeomorphism, and so does not provide a one-to-one mapping between points on $TQ$ and points on $T^*Q$. It is for this reason that the $\lambda_\alpha$ are included. The origin of the secondary (and all further) constraints is dynamical - they are the conditions which must be satisfied in order for the dynamics to be consistent. As such, it is not necessary to introduce additional Lagrange multipliers into the Hamiltonian.\\

Once the algorithm terminates, the full collection of constraints obtained may be separated into first and second-class functions exactly as described in section (\ref{Subsec:Geometry}). In addition, we may follow the steps leading to and including the introduction of the Dirac bracket. Recall that the utility of this structure lies in the fact that it allows the second-class constraints to be imposed as strong equalities before computing the equations of motion. 
\section{An Example}\label{Sec:Eg1}
In an effort to make the content of the preceding sections somewhat less abstract, we shall consider an example of an almost-regular system, which allows each of the steps of the geometrical constraint procedure to be appreciated in a simple setting. Of course, our choice of example is motivated entirely by pedagogical considerations, and should therefore not be considered to reflect the dynamics of a genuine physical system.\\

For this example, we shall work with the configuration space $Q=\mathbb{R}^4$, with local coordinates $(q_1,q_2,q_3,q_4)$. Our choice of Lagrangian function is
\begin{equation}\label{Eq:Eg1Lag}
    L(q,\dot q)=\frac{1}{2} \left[ \left(q_1 + \dot {q}_2 + \dot{q}_3\right)^2 + \left(\dot{q}_4-\dot{q}_2\right)^2 - 2q_2q_4 \right]
\end{equation}
The Hessian matrix $W_{ij}$ with respect to the velocities $\dot{q}_i$ is readily calculated to be
\begin{equation*}
    W_{ij} = \begin{pmatrix}
        0 & 0 & 0 & 0\\
        0 & 2 & 1 & -1\\
        0 & 1 & 1 & 0\\
        0 & -1 & 0 & 1
    \end{pmatrix}
\end{equation*}
so that $\textrm{rank}\,W_{ij}=2$, which is non-maximal. The system is thus singular, and we expect the appearance of constraints. Indeed, the momenta canonically conjugate to the $\dot{q}_i$ are
\begin{equation}\label{Eq:Eg1Momenta}
    p_1 =0\quad\quad p_2=q_1+\dot{q}_2+\dot{q}_3 -\left(\dot{q}_4-\dot{q}_2\right) \quad\quad p_3 = q_1+\dot{q}_2+\dot{q}_3 \quad\quad p_4 = \dot{q}_4-\dot{q}_2
\end{equation}
From this, we immediately see that there appear two primary constraints
\begin{equation}\label{Eq:Eg1PrimaryConstraints}
    \phi^1 := p_1 \hspace{1.8cm} \phi^2:= p_2-p_3+p_4
\end{equation}
The zero-set of $\phi^1$ and $\phi^2$ defines the primary constraint manifold $M_0\subset T^*\mathbb{R}^4\cong \mathbb{R}^8$
\begin{equation}\label{Eq:Eg1M0}
    M_0:= \{\,x\in Q\;|\; \phi^1(x) = 0\;;\;\phi^2(x)=0\,\}\;\subset\;T^*Q
\end{equation}
Following the steps of section (\ref{Sec:Pre-symplecticAlgorithm}), we turn to the Hamiltonian function $H_0$, obtained as a direct Legendre transform of $L$. From the momenta (\ref{Eq:Eg1Momenta}), it is not too hard to see that
\begin{equation}\label{Eq:Eg1H0}
    H_0 = \frac{1}{2}\left[ p_3^2+p_4^2 - 2q_1p_3 + 2q_2q_4 \right]
\end{equation}
The restriction of the canonical symplectic form $\omega$ to $M_0$ reads
\begin{equation}\label{Eq:Eg1omega0}
    \omega_0=dq_2 \wedge dp_2 + dq_3\wedge dp_3 + dq_4\wedge\left(dp_3-dp_2\right) \quad\quad\implies\quad\quad\textrm{ker}\,\omega_0 \,= \; \biggr\langle \frac{\partial}{\partial q_1}\;,\,\frac{\partial}{\partial q_2} - \frac{\partial}{\partial q_3}+\frac{\partial}{\partial q_4} \biggr\rangle
\end{equation}
where the angle brackets refer to the span of the vector fields. From (\ref{Eq:Eg1H0}) and (\ref{Eq:Eg1omega0}), it follows that requiring $\langle dH_0,\textrm{ker}\,\omega_0\rangle=0$ gives a pair of secondary constraints
\begin{equation}\label{Eq:Eg1Secondary}
    \langle dH_0,\textrm{ker}\,\omega_0\rangle=0 \quad\quad\implies\quad\quad \phi^3:= p_3 \approx0, \quad\quad \phi^4:= q_2+q_4 \approx0
\end{equation}
The zero-set of the four constraint functions $\phi^{\alpha}$ defines the secondary constraint submanifold $\jmath_1:M_1\hookrightarrow M_0$. In order to deduce whether the algorithm produces further constraints, we must calculate the symplectic orthogonal $TM_1^{\perp}$. The most expedient way to do so is to note that if we consider a general vector $X\in T(T^*Q)$, expressed locally as $X=A_i\partial_{q_i}+B_i\partial_{p_i}$, then $X$ may only be tangent to the surface $M_1$ if $X[\phi^{\alpha}]=0$ for $\alpha=1,2,3,4$. It is then straightforward to verify that $TM_1$ is spanned by the following vectors
\begin{equation*}
    TM_1 \,=\;\; \biggr\langle \frac{\partial}{\partial q_1}\,,\, \frac{\partial}{\partial q_2}-\frac{\partial}{\partial q_4}\,,\, \frac{\partial}{\partial q_3} \; , \; \frac{\partial}{\partial p_2} - \frac{\partial}{\partial p_4}\,\biggr\rangle
\end{equation*}
Restricting the canonical symplectic form $\omega$ to $M_1$ yields $\omega_1=2\, dq_2\wedge dp_2$, which together with the basis given above for $TM_1$, allows us to deduce that
\begin{equation}
    TM_1^{\perp} \,= \;\;\biggr\langle \frac{\partial}{\partial q_1}\;,\; \frac{\partial}{\partial q_3}\,\biggr\rangle
\end{equation}
It is then relatively clear that no new constraints arise by imposing that $\langle dH_0,TM_1^{\perp}\rangle=0$. We simply recover $\phi^3=p_3\approx0$, and the algorithm stabilises on $M_1$. We therefore declare
\begin{equation}\label{Eq:Eg1M1}
    M_1 = \{\,x\in Q\;|\; \phi^1(x) = 0\;;\;\phi^2(x)=0\;;\;\phi^3(x)=0\;;\;\phi^4(x)=0\,\}:= M_f
\end{equation}
From our four constraints
\begin{equation}\label{Eq:Eg13Constraints}
    \phi^1=p_1 \hspace{1.4cm} \phi^2= p_2-p_3+p_4 \hspace{1.4cm} \phi^3 = p_3 \hspace{1.4cm} \phi^4= q_2+q_4
\end{equation}
we wish to compute the matrix $J^{\alpha\beta}:=\{\phi^{\alpha},\phi^{\beta}\}$ introduced in section (\ref{Subsec:Geometry}). From the local expression (\ref{Eq:LocalPB}), it is straightforward to deduce that
\begin{equation}\label{Eq:Eg1JMatrix}
    J = \begin{pmatrix}
        0 & 0 & 0 & 0\\
        0 & 0 & 0 & -2\\
        0 & 0 & 0 & 0\\
        0 & 2 & 0 & 0
    \end{pmatrix}\hspace{1.6cm} \textrm{rank}\,J=2
\end{equation}
The rank of this $(4\times 4)$ matrix is two, and so we expect to be able to extract two $C^{\infty}$-linear combinations of the $\phi^{\alpha}$ which are first-class. However, the calculational machinery of computing null eigenvectors of $J$ is somewhat overkill here - it is quite clear that both $\phi^1$ and $\phi^3$ weakly Poisson-commute with all constraints, and so are first-class. In accordance with the notation of (\ref{Subsec:Geometry}), we therefore write
\begin{equation}\label{Eq:Eg1FirstClassOmega}
    \Omega^1:=p_1 \hspace{1.4cm} \Omega^2:=p_3 \hspace{1.4cm} \chi^1 := p_2-p_3+p_4 \hspace{1.4cm} \chi^2:= q_2+q_4
\end{equation}
The matrix $C^{ij}:=\{\chi^i,\chi^j\}$ is clearly invertible, since we have
\begin{equation*}
    C=\begin{pmatrix}
        0 & -2\\
        2 & 0
    \end{pmatrix}\quad\quad\implies \quad\quad C^{-1} = \frac{1}{2}\begin{pmatrix}
        0 & 1\\
        -1 & 0
    \end{pmatrix}
\end{equation*}
It thus follows that the Dirac bracket between two phase space functions $f,g\in C^{\infty}(T^*Q)$ reads
\begin{equation}\label{Eq:Eg1DiracBracket}
    \{f,g\}_{D}=\{f,g\} -\frac{1}{2}\{f,\chi^1\}\{\chi^2,g\} + \frac{1}{2} \{f,\chi^2\} \{\chi^1,g\}
\end{equation}
The equations of motion on the constraint surface $M_f$ may now be calculated in a straightforward manner. We begin by imposing the second-class constraints as strong equalities. Note that this involves some degree of choice, and referring to the discussion and notation of section (\ref{Subsec:Geometry}), corresponds to choosing a set of local coordinates on $M_{\chi}$. For our particular example, we choose to work with $(q_1,q_3,q_4,p_1,p_3,p_4)$. As described previously, $M_{\chi}$ possesses a closed 2-form $\omega_{\chi}$ which is non-degenerate. The introduction of the Dirac bracket is what allows us to encode the symplectic structure of $M_\chi$ directly on $T^*Q$. Imposing $\chi^1$ and $\chi^2$ as strong equalities, the resulting function $H$ is then formally extended from $M_{\chi}$ to $T^*Q$. Note that this extension is not unique, as the Dirac bracket ensures that the dynamics is insensitive to our particular choice. The sole requirement is that when restricted to $M_{\chi}$, the extended function must coincide with $H$. For simplicity, therefore, we choose to use the same coordinate representation for both functions. Finally, we must add the product of an undetermined Lagrange multiplier $\lambda_a$ and a first-class constraint $\Omega^a$, for $a=1,2$. Thus, in total, we have
\begin{equation}\label{Eq:Eg1HT}
    H_T= \frac{1}{2}\left(p_3^2 + p_4^2 - 2q_1 p_3 - 2 q_4^2 \right)+ \lambda_1p_1 + \lambda_2p_3
\end{equation}
The most general dynamical evolution - including gauge freedom - of any phase space variable is then calculated via its Dirac bracket with the total Hamiltonian. We find that
\begin{align*}
    \dot{q}_1 &= \lambda_1 & \dot{q}_3 &= p_3-q_1 +\frac{1}{2}p_4+\lambda_2 \approx-q_1 +\frac{1}{2}p_4+\lambda_2 & \dot{q}_4 &= \frac{1}{2}p_4\\
    \dot{p}_1&=p_3\approx0  & \dot{p}_3&=0 & \dot{p}_4&=q_4
\end{align*}
The dynamical evolution of $(q_2,p_2)$ is then deduced from the second-class constraints. With this example, we conclude our study of singular symplectic systems. In the sections that follow, we provide an introduction to contact and pre-contact geometry, as well as a constraint algorithm which parallels that of section (\ref{Sec:Pre-symplecticAlgorithm}).
\section{Contact Geometry}\label{Sec:ContactGeometry}
As described in the introduction, contact structures arise naturally when scaling degrees of freedom are eliminated from the ontology of theories of mechanical systems. In general, a contact structure on a smooth manifold $C$, of dimension $2n+1$, is a maximally non-integrable distribution $\xi\subset TC$. Locally, this distribution may be described as the kernel of some $\eta\in\Omega^1(U\subset C)$, referred to as a contact form \cite{bravetti2019contact,de_Le_n_2019,de2020review,de_Le_n_2020,de2019contact}. If the quotient line bundle $TC/\xi \rightarrow C$ is trivial, as we shall assume to be the case, then $\xi$ is said to be \textbf{coorientable}, and $\eta$ may be extended to a global contact form on $C$. With the assumption of coorientability, we henceforth designate a contact manifold via $(C,\eta)$; every such manifold admits a distinguished \textbf{Reeb vector field} $\mathcal{R}\in\mathfrak{X}^{\infty}(C)$, defined via
\begin{equation}\label{Eq:Reeb}
    \iota_{\mathcal{R}}d\eta = 0 \quad\quad\quad\quad \iota_{\mathcal{R}}\eta =1
\end{equation}
As for symplectic manifolds, around each point $p\in C$, we may always find a local chart of Darboux coordinates $(x^1,\,\cdots,x^n,y_1,\,\cdots,y_n,z)$, in which $\eta$ takes the form $\eta = dz - y_i\,dx^i$. We shall refer to this as the \textbf{canonical contact form}.\\

The extended tangent bundle $TQ\times \mathbb{R}$ of the $n$-dimensional configuration space $Q$ is a manifold of dimension $2n+1$, with local coordinates $(q^i,v^i,z)$. A contact Lagrangian is a function $L:TQ\times\mathbb{R}\rightarrow \mathbb{R}$, from which we define the Cartan forms $\theta_L$ and $\omega_L$, together with the energy function $E_L$. This procedure goes through exactly as on $TQ$, as the canonical structures introduced in section (\ref{Sec:GeometricPreliminaries}) have natural extensions to $TQ\times \mathbb{R}$. The space $TQ\times\mathbb{R}$ is made into a (pre-)contact manifold, introducing the 1-form $\eta_L:=dz-\theta_L$, and we refer to the triple $(TQ\times\mathbb{R},\eta_L,E_L)$ as a (pre-)contact Lagrangian system. If the characteristic distribution $\mathcal{C}:=\textrm{ker}\,\eta_L\,\cap\,\textrm{ker}\,d\eta_L$ is of rank $2(n-k)$, we say that $\eta_L$ is \textbf{of class} $\boldsymbol{2k+1}$; locally, this implies
\begin{equation}\label{Eq:Class}
    \eta\wedge d\eta^k\neq0\quad\quad\textrm{but}\quad\quad \eta\wedge d\eta^{k+1}=0
\end{equation}
Clearly, when $\textrm{rank}\;\mathcal{C}=2n$, $(TQ\times\mathbb{R},\eta_L,E_L)$ is a regular system; in such a case, the Legendre map $FL:TQ\times \mathbb{R}\rightarrow T^*Q\times\mathbb{R}$ is a diffeomorphism, and acts on local coordinates $(q^i,p_i,z)$ of $T^*Q\times\mathbb{R}$ as
\begin{equation}
    FL^* q^i=q^i\quad\quad FL^*p_i = \frac{\partial L}{\partial v^i}\quad\quad FL^* z = z
\end{equation}
The space $T^*Q\times\mathbb{R}$ has a canonical contact form $\eta= dz-p_i\,dq^i$, and the Reeb vector field $\mathcal{R}$ follows from (\ref{Eq:Reeb}). For regular systems, we take $\mathcal{R}=\partial/\partial z$; however, when the manifold is endowed with a pre-contact form, the Reeb field ceases to be unique.\\

If $(TQ\times\mathbb{R},\eta_L,E_L)$ is a hyperregular Lagrangian system (that is to say, one for which $FL$ is a global diffeomorphism) we introduce the unique function $H:T^*Q\times\mathbb{R}\rightarrow\mathbb{R}$, such that $FL^*H=E_L$. The triple $(T^*Q\times\mathbb{R},\eta,H)$ is then said to constitute a contact Hamiltonian system, and there exists a bundle morphism
\begin{equation}\label{Eq:bHmorphism}
    \begin{split}
        \bar{\flat}\;:\; T(T^*Q\times\mathbb{R}) &\longrightarrow T^*(T^*Q\times \mathbb{R})\\
        v\;&\longmapsto \iota_v\,d\eta + (\iota_v\eta)\eta
    \end{split}
\end{equation}
The equations of motion are deduced seeking a vector field $X_{H}\in \mathfrak{X}^{\infty}(T^*Q\times\mathbb{R})$, which satisfies
\begin{equation}\label{Eq:GeometricalHamiltonEquations}
    \bar{\flat}(X_H) = dH - \left( \mathcal{R}(H)+H\right)\eta
\end{equation}
Suppose that we decompose the vector field $X_H$ as
\begin{equation}\label{Eq:XH}
    X_H = A^i\frac{\partial}{\partial q^i} + B_i\frac{\partial}{\partial p_i} + C\frac{\partial}{\partial z}
\end{equation}
then the coefficient functions $A^i$, $B_i$, and $C$ satisfy
\begin{equation}\label{Eq:LocalHamiltonEquations}
    A^i = \frac{\partial H}{\partial p_i} \quad\quad B_i = - \,\biggr(\frac{\partial H}{\partial q^i} + p_i \,\frac{\partial H}{\partial z}\biggr) \quad\quad C = p_i\, \frac{\partial H}{\partial p_i} - H
\end{equation}
\section{Pre-contact Constraint Algorithm}\label{Sec:PreContactConstraintAlgorithm}
For the geometrical constraint procedure, we shall suppose that $(TQ\times\mathbb{R},\eta_L,E_L)$ is a pre-contact system, with $\eta_L$ being of class $2k+1$. In practice, this latter condition is most easily verified noting that when $\eta_L$ is of class $2k+1$, the Hessian matrix with respect to the velocities $v^i$ is of rank $k$. The image $P_0:=FL(TQ\times\mathbb{R})$ is assumed to be a closed submanifold of $T^*Q\times\mathbb{R}$, with inclusion $\kappa_0:P_0\hookrightarrow T^*Q\times\mathbb{R}$. As for the pre-symplectic algorithm $P_0$ defines the primary constraint surface, and the canonical contact form $\eta$ on $T^*Q\times\mathbb{R}$ is restricted to $\eta_0=\kappa_0^*\eta$. The pre-contact structure induced on $P_0$ by $\eta_0$ is then made into a Hamiltonian system by introducing the function $H_0:P_0\rightarrow \mathbb{R}$, which satisfies $FL_0^*\,H_0=E_L$, where $FL_0$ is the restriction of $FL$ to its image. Finally, we restrict the bundle morphism $\bar{\flat}$ to $P_0$, denoting the resulting map $\bar{\flat}_0$.\\

With these notational conventions established, we may introduce a notion of orthogonality, which parallels the symplectic orthogonal (\ref{Eq:SymplecticOrthogonal}). Note, however, that since $\bar{\flat}_0(X)(Y)$ is neither symmetric nor antisymmetric under the exchange of its arguments, for a given distribution $\mathcal{D}\subset TP_0$, there exists a concept of both left and right orthogonality, denoted $^{\perp}\mathcal{D}$ and $\mathcal{D}^{\perp}$ respectively. We shall only be concerned with right orthogonality, and so define
\begin{equation}\label{Eq:DOrthogonal}
    \mathcal{D}^{\perp}:=\{v\in TP_0\;|\; \bar{\flat}_0(w)(v)=0\quad\textrm{for all}\; w\in \mathcal{D}\,\}
\end{equation}
In the interest of reducing cumbersome notation in what follows, we introduce
\begin{equation*}
     \alpha_0:=dH_0 - \left(\mathcal{R}(H_0)+H_0\right)\eta_0
\end{equation*}
so that the dynamical problem to be solved is that of finding a vector field $X_H\in\mathfrak{X}^{\infty}(P_0)$, such that
\begin{equation}\label{Eq:ContactHamiltonianEquation}
    \bar{\flat}_0(X_H) = \alpha_0
\end{equation}
The first step of the constraint procedure requires that we restrict $P_0$ to the subset of points at which $\alpha_0$ is in the range of $\bar{\flat}_0$: that is to say
\begin{equation}\label{Eq:P1}
    P_1:=\{x\in P_0\;|\; (\alpha_0)_x\in \bar{\flat}_0(T_xP_0)\,\}
\end{equation}
As was the case for the pre-symplectic constraint algorithm, it is convenient to make use of the relationship $(TP_0^{\perp})^o= \bar{\flat}_0(TP_0)$. Further, it is found that $TP_0^{\perp}$ is precisely the characteristic distribution $\mathcal{C}$, which provides the following, more practical condition for deducing $P_1$
\begin{equation}\label{Eq:P1Again}
    P_1=\{x\in P_0\;|\;\langle (\alpha_0)_x,\mathcal{C}_x\rangle=0\,\}
\end{equation}
We must now impose that any vector field solution on $P_1$ be tangent to this space, which generally requires us to restrict to the smaller set
\begin{equation}\label{Eq:P2}
    P_2:=\{x\in P_1\;|\; \langle (\alpha_0)_x,T_xP_1^{\perp}\rangle=0\,\}
\end{equation}
The tangency condition is then imposed recursively, producing a series of embedded submanifolds $\kappa_{i+1}:P_{i+1}\hookrightarrow P_i$, until we reach the final constraint manifold $P_f$, where further iteration of the algorithm produces no new conditions. In this case, the geometric equation $\bar{\flat}_0(X_H)=\alpha_0$ has a well-defined, albeit non-unique solution which is everywhere-tangent to $P_f$. The non-uniqueness follows from the observation that we may add to $X_H$ any element of $\mathcal{C}\,\cap \,TP_f$, without altering the physical dynamics or leaving the constraint surface.\\

In order to present the complementary local description, we require a bracket structure appropriate for the odd-dimensional spaces of contact geometry. This is provided by the \textbf{Jacobi bracket}, where, for two functions $f,g\in C^{\infty}(T^*Q\times\mathbb{R})$, we have
\begin{equation}\label{Eq:LocalJacobiBracket}
    \{f,g\}_J = \frac{\partial f}{\partial q^i}\frac{\partial g}{\partial p_i} - \frac{\partial f}{\partial p_i}\frac{\partial g}{\partial q^i} + \left(p_i\frac{\partial g}{\partial p_i}- g\right)\frac{\partial f}{\partial z} - \left( p_i\frac{\partial f}{\partial p_i}-f\right)\frac{\partial g}{\partial z}
\end{equation}
Note that the pair $\left(C^{\infty}(T^*Q\times\mathbb{R}) , \{\,\cdot \;,\,\cdot \,\}_J\right)$ constitutes a Lie algebra; however the Jacobi bracket satisfies only a weaker form of the Leibniz rule. In particular, for three functions $f,g,h\in C^{\infty}(T^*Q\times \mathbb{R})$, we have
\begin{equation}\label{Eq:GeneralisedLeibniz}
    \{f,gh\}_J = g\,\{f,h\}+ h\,\{f,g\}_J + gh\mathcal{R}(f)
\end{equation}
Suppose that $P_f$ may be described locally as the zero-set of the $N$ functions $\{\Phi^{\alpha}\}$. Introducing the matrix of brackets $K^{\alpha\beta}:=\{\Phi^{\alpha},\Phi^{\beta}\}_J$, with $\textrm{rank}\,K=N-F$, we extract $F$ first-class functions $\Omega^a$, and $N-F$ second-class functions $\chi^j$, following precisely the same procedure as in (\ref{Eq:FirstClassv}). The Jacobi bracket of the second-class constraints again forms an invertible matrix $C^{ij}:=\{\chi^i,\chi^j\}_J$ of size $(N-F)\times (N-F)$, whose inverse we denote $C_{ij}$. In close analogy to the symplectic case, the second-class constraints are imposed as strong equalities, and the Jacobi bracket replaced with the \textbf{Dirac-Jacobi bracket} $\{\,\cdot\;,\,\cdot\,\}_{DJ}$. For two functions $f,g\in C^{\infty}(T^*Q\times\mathbb{R})$, we have
\begin{equation}\label{Eq:DiracJacobiBracket}
    \{f,g\}_{DJ}:=\{f,g\}_J - \{f,\chi^i\}_J\,C_{ij}\{\chi^j,g\}_J
\end{equation}
The time evolution of a function $f$ on $T^*Q\times\mathbb{R}$ is not deduced - as was the case for symplectic systems - via its Dirac-Jacobi bracket with the total Hamiltonian. Instead, we are required to introduce additional structure. Until now, our treatment of contact manifolds has been relatively low-key, favouring practicality over formality. However, at present, it is of benefit to digress slightly into a more technical discussion. Specifically, we highlight that contact manifolds are a particular case of a more general class of objects known as \textbf{Jacobi manifolds} \cite{de1997geometric}. Such manifolds are characterised by a pair $(\Lambda,E)$, where $\Lambda$ is a skew-symmetric contravariant tensor field of rank two, and $E$ a vector field. For a contact manifold, $E$ is simply the Reeb field $\mathcal{R}$, and $\Lambda$ has the following local coordinate expression
\begin{equation}\label{Eq:LocalLambda}
    \Lambda = \frac{\partial}{\partial q^i}\wedge\frac{\partial}{\partial p_i} + p_i\,\frac{\partial}{\partial z}\wedge \frac{\partial}{\partial p_i}
\end{equation}
This bivector field induces a map $\sharp_{\Lambda}$ between covectors and vectors, whose action on some covector $\xi$ is $\sharp_{\Lambda}(\xi):=\Lambda(\,\cdot\;,\xi)$. Indeed, the local expression (\ref{Eq:LocalJacobiBracket}) may be expressed somewhat more compactly as
\begin{equation}
    \{f,g\}_J=\Lambda(df,dg) + f\mathcal{R}(g)-g\mathcal{R}(f)
\end{equation}
In addition to $(\Lambda,\mathcal{R})$, the extended cotangent bundle admits a second Jacobi structure $(\Lambda_{DJ},\mathcal{R}_{DJ})$, where
\begin{equation}\label{Eq:RDJ}
    \mathcal{R}_{DJ}:= \mathcal{R} + C_{ij}\mathcal{R}(\chi^j)\biggr[\chi^i\mathcal{R}-\sharp_{\Lambda}(d\chi^i)\biggr]
\end{equation}
Having imposed the second-class constraints as strong equalities, we must work with the Jacobi structure $(\Lambda_{DJ},\mathcal{R}_{DJ})$ for subsequent calculations. Such is the case, for example, when computing the contact Hamiltonian vector field $X_f$ corresponding to a function $f$, where we have
\begin{equation}\label{Eq:LambdaVectorField}
    X_f:= \sharp_{\Lambda_{DJ}}(df) - f\mathcal{R}_{DJ}
\end{equation}
Finally, we introduce the total Hamiltonian $H_T:=H+u_a\Omega^a$, in which $H$ is an arbitrary extension of $H_0$ to $T^*Q\times\mathbb{R}$, and $u_a$ are undetermined Lagrange multipliers. The time evolution of a function $g\in C^{\infty}(T^*Q\times\mathbb{R})$ is then given by
\begin{equation}\label{Eq:Evolution}
    \dot{g}=X_{H_T}(g) = \{g,H_T\}_{DJ} -g\mathcal{R}_{DJ}(H_T)
\end{equation}
When evaluated on the constraint surface $P_f$, the first-class constraints $\Omega^a$ vanish identically, leaving
\begin{equation}\label{Eq:fEvolution2}
    \dot{g}\approx \left(X_H + u_aX_{\Omega^a}\right)(g)
\end{equation}
This expression allows us to easily deduce the evolution of any function on the (extended) phase space along the Hamiltonian flow. Having concluded our study of pre-contact geometry and the geometrical constraint algorithm for pre-contact Hamiltonians, we close with an example, mirroring that of section (\ref{Sec:Eg1}).
\section{An Example}\label{Sec:Eg2}
For our concluding example, we shall work over the two-dimensional configuration space $Q=\mathbb{R}^2$, parameterised via the local coordinates $(q_1,q_2)$. The Lagrangian function on the extended tangent bundle $TQ\times\mathbb{R}\cong \mathbb{R}^5$ has the following form
\begin{equation}\label{Eq:Eg2Lag}
    L(q,\dot{q},z) = \frac{1}{2}\left(\dot{q}_1+\dot{q}_2\right)^2  +q_1 +q_2z
\end{equation}
The Hessian matrix with respect to $(\dot{q}_1,\dot{q}_2)$ reads
\begin{equation*}
    W=\begin{pmatrix}
        1 & 1\\
        1 & 1
    \end{pmatrix}
\end{equation*}
which is clearly of rank one. This means that the 1-form $\eta_L$, given locally by
\begin{equation*}
    \eta_L = dz - \left(\dot{q}_1+\dot{q}_2\right)\left(dq_1+dq_2\right)
\end{equation*}
is a pre-contact form of class three. Indeed, we note that this may also be deduced from the observation that $\eta_L\,\wedge \,d\eta_L\neq0$, but $\eta_L\,\wedge\, (d\eta_L)^2=0$. The momenta $p_i$ associated with the $\dot{q}_i$ are given by $p_1 = \dot{q}_1+\dot{q}_2=p_2$, and so we obtain the primary constraint $\Phi^1:= p_1-p_2$, the zero-set of which defines the submanifold $P_0\hookrightarrow T^*Q\times\mathbb{R}$. The canonical contact form $\eta$ on $T^*Q\times\mathbb{R}$ restricts to the pre-contact form $\eta_0=dz-p_1\left(dq_1+dq_2\right)$, and the Hamiltonian function $H_0:P_0\rightarrow \mathbb{R}$ adopts the form
\begin{equation}\label{Eq:Eg2H0}
    H_0 = \frac{1}{2}p_1^2-q_1-q_2z
\end{equation}
The first step of the constraint algorithm requires that we deduce the characteristic distribution $\mathcal{C}$ of $\eta_0$; recalling that $\mathcal{C}:=\textrm{ker}\,\eta_0\cap\textrm{ker}\,d\eta_0$, it is relatively straightforward to see that
\begin{equation}\label{Eq:Eg2Characteristic}
    \mathcal{C}\;=\;\,\biggr\langle \,\frac{\partial}{\partial q_1}- \frac{\partial}{\partial q_2}\,\biggr\rangle
\end{equation}
Taking the Reeb field to be in the $z$ direction - $\mathcal{R} =\partial/\partial z$ - the 1-form $\alpha_0:=dH_0 - \left(\mathcal{R}(H_0)+H_0\right)\eta_0$ reads
\begin{equation*}
    \alpha_0 = p_1\,dp_1-dq_1 - z\,dq_2-q_2\,dz -\left[\frac{1}{2}p_1^2-q_1-q_2(1+z)\right]\eta_0
\end{equation*}
Referring to the characterisation (\ref{Eq:P1Again}) of the secondary submanifold $P_1$, we see that demanding $\langle\alpha_0,\mathcal{C}\rangle=0$ produces the constraint $\Phi^2:=1-z$, and so $P_1$ is the zero-set of both $\Phi^1$ and $\Phi^2$. The tangent space $TP_1$ is spanned locally by the following vectors
\begin{equation}\label{Eq:Eg2TP1}
    TP_1 \;=\;\,\biggr\langle\, \frac{\partial}{\partial q_1}\,,\, \frac{\partial}{\partial q_2}\,,\,\frac{\partial}{\partial p_1}+\frac{\partial}{\partial p_2}\,\biggr\rangle
\end{equation}
The orthogonal space $TP_1^{\perp}$ is then deduced by demanding that a general vector $X\in TP_1$ satisfy the condition $\iota_Xd\eta_1 + \left(\iota_X \eta_1\right)\eta_1=0$, in which $\eta_1$ denotes the restriction of $\eta$ to $P_1$. From (\ref{Eq:Eg2TP1}), the vector $X$ may be expressed locally as $X = A\,\partial_{q_1} + B\,\partial_{q_2} + C\left(\partial_{p_1}+\partial_{p_2}\right)$; we then have
\begin{equation*}
    \iota_Xd\eta_1 + \left(\iota_X \eta_1\right)\eta_1= \left(A+B\right)dp_1 +\left(p_1^2\left(A+B\right)-C\right)\left(dq_1+dq_2\right)\stackrel{!}{=}0
\end{equation*}
This fixes $A+B=0$, and $C=0$, so that
\begin{equation*}
    TP_1^{\perp} \;=\;\, \biggr\langle \,\frac{\partial}{\partial q_1}-\frac{\partial}{\partial q_2}\,\biggr\rangle\; = \; TP_0^{\perp}
\end{equation*}
From this, it is clear that the algorithm must stabilise on $P_1$, and we have
\begin{equation}\label{Eq:Eg2Pf}
    P_1=\{x\in T^*Q\times\mathbb{R}\;|\; \Phi^1(x) = p_1-p_2 =0\;\,;\;\Phi^2(x) = 1-z=0\,\}\,:=\,P_f
\end{equation}
Since the algorithm terminates after a single iteration, we should classify the constraints $\Phi^1$ and $\Phi^2$, for which, in section (\ref{Sec:PreContactConstraintAlgorithm}), we introduced the matrix $K^{\alpha\beta}:=\{\Phi^{\alpha},\Phi^{\beta}\}_J$. From (\ref{Eq:LocalJacobiBracket}), it is easy to see that $K$ is simply the zero matrix, with which we conclude that both constraints are first-class, and that the Dirac-Jacobi bracket therefore coincides with the Jacobi bracket.\\

The total Hamiltonian reads
\begin{equation}\label{Eq:Eg2HT}
    H_T = \frac{1}{2}p_1^2-q_1-q_2z + u_1\left(p_1-p_2\right) + u_2\left(1-z\right)
\end{equation}
in which $u_1$ and $u_2$ are arbitrary Lagrange multipliers. The time evolution of a general phase space function $f$ follows from
\begin{equation}
    \dot{f} = \{f,H_T\} + \left(\frac{1}{2}p_1^2 +q_1+q_2z - u_2\left(1-z\right)\right) \frac{\partial f}{\partial z} + \left(q_2+u_2\right) p_i\frac{\partial f}{\partial p_i}
\end{equation}
Note that the quantity $\{f,H_T\}$ on the RHS of this expression refers to the \textit{Poisson} bracket. For example, we find that
\begin{equation*}
    \dot{q}_1 = p_1 - q_1\left(q_2+u_2\right) + u_1 \hspace{1.2cm} \dot{z}=\frac{1}{2}p_1^2 +q_1+u_1\left(1-2z\right) \approx \frac{1}{2}p_1^2 +q_1-u_1
\end{equation*}
We leave it as an exercise to compute the remaining Jacobi brackets.
\section{Concluding Remarks}\label{Sec:Conc}
Singular systems play a prominent role in a large number of physical contexts. In this article, we have provided a review of the fundamental geometrical concepts required to describe such systems. In addition to our presentation of pre-symplectic and pre-contact geometry, together with their implementation in the description of mechanical systems, we have also shown how one can use geometrical principles to systematically deduce the subsets of phase space upon which a system's dynamical equations admit well-defined solutions. We have seen that constraints are broadly categorised into two types: first and second-class. Our characterisation of a constraint as being `first-class' was based upon its bracket (Poisson or Jacobi) with all other constraints vanishing on the final constraint surface. Typically, first-class constraints are attributed to gauge degrees of freedom; however, as discussed, this so-called Dirac conjecture - that all first-class constraints are generators of gauge transformations - is not true in full generality. A more detailed and particularly readable account of these ideas may be found in \cite{cabo1986dirac}.
\bibliographystyle{unsrt}
\bibliography{Refs}
\end{document}